\documentclass{emulateapj}
\usepackage{epsf}
\usepackage{bm,color}

\def\gsim{\mathrel{\rlap{\lower 4pt \hbox{\hskip 1pt $\sim$}}\raise 1pt
\hbox {$>$}}}
\def\lsim{\mathrel{\rlap{\lower 4pt \hbox{\hskip 1pt $\sim$}}\raise 1pt
\hbox {$<$}}}

\newcommand{\bra}{<\hspace{-4pt}}
\newcommand{\ket}{\hspace{-4pt}>}
\begin{document}
\title{Dissipation of magnetic fields in star-forming clouds with different metallicities}
\author{Hajime Susa$^1$}
\author{Kentaro Doi$^{1}$}
\author{Kazuyuki Omukai$^2$}
\affil{Department of Physics, Konan University, Okamoto, Kobe, Japan\altaffilmark{1}}
\affil{Astronomical Institute, Tohoku University, Aramaki, Sendai, Japan\altaffilmark{2}}
\email{susa@konan-u.ac.jp}
\begin{abstract}
We study dissipation process of magnetic fields in the metallicity range $0-1 Z_{\odot}$ for  
contracting prestellar cloud cores. 
By solving non-equilibrium chemistry for important charged species including charged grains, 
we evaluate the drift velocity of the magnetic-field lines with
respect to the gas. 
We find that the magnetic flux dissipates in the density range 
$10^{12}{\rm cm^{-3}} \la n_{\rm H} \la 10^{17}{\rm cm^{-3}}$ for the solar-metallicity case
at the scale of the core, which is assumed to be the Jeans scale.
The dissipation density range becomes narrower for lower metallicity. 
The magnetic field is always frozen to the gas below metallicity $\la 10^{-7}-10^{-6}Z_\odot$, depending 
on the ionization rate by cosmic rays and/or radioactivity. 
With the same metallicity, the dissipation density range becomes wider for lower ionization rate.
The presence of such a dissipative regime is expected to 
cause various dynamical phenomena in protostellar evolution
such as the suppression of jet/outflow launching and fragmentation of the circumstellar disks
depending on the metallicity.


\end{abstract}
\keywords{early Universe---metal-poor stars}
\section{Introduction}
The first episode of star formation makes a large impact on the subsequent thermal and 
chemical evolution of the universe by initiating such events as metal enrichment 
and reionization of the intergalactic medium. 
The first stars, so-called population III stars, are formed from the primordial pristine gas, 
consisting only of hydrogen, helium and a trace amount of deuterium and lithium.
Theoretical studies predict that they are typically very massive $\sim 100M_{\odot}$ 
\citep[e.g.][]{bromm02, omukai03, mckee08, hosokawa11,susa13,hirano14,susa14}, and some of them end their lives 
as supernova explosions after a few million years \citep{umeda02, heg02}.

Following the explosions, metal-enriched ejecta is mixed with the ambient material 
and may recollapse to form the next generation of stars (e.g., Ritter et al. 2012). 
Metallicity of the newly formed stars reflects the degree of mixing of the ejecta with 
the pristine gas \citep{wise12,chen14}.
A gas with metals cools more efficiently owing to the higher radiation emissivity 
as well as higher sustainability against stellar radiative feedback.  
Since the cold environment thus achieved is favorable for gravitational collapse and fragmentation 
of star-forming clouds, vigorous star formation might be induced by the metal enrichment.
In fact, in some numerical models \citep{ahn12,wise14}, 
second-generation stars formed from metal-enriched gas, rather than 
the first stars, are the dominant radiation sources for the cosmic reionization. 
Knowledge of star formation in low-metallicity environments is a clue to understand 
the cosmic structure formation.

Thermal and dynamical properties of low-metallicity star-forming clouds has been
investigated by a number of semi-analytic \citep{omukai00,schneider02,omukai05,omukai10,ji14} 
and numerical studies \citep{bromm01,machida08_Z,clark08,dopcke11,dopcke13}.
With slight metal enrichment of $10^{-6}-10^{-4}Z_\odot$, 
the cooling by dust grains causes rapid cooling at high enough density ($>10^{10}{\rm cm^{-3}}$)
where the Jeans mass is small ($\la 1 M_{\odot}$), thereby triggering fragmentation of the clouds 
into small pieces \citep{omukai00,schneider02,schneider03,omukai05,schneider06,omukai10,schneider12}. 
At this critical metallicity, whose exact value depends on the nature and 
amount of dust in the gas, the transition of characteristic stellar mass 
from massive to low-mass ones is expected to occur \citep[but see][]{frebel07}. 

Most studies so far on low-metallicity star formation have not taken account of 
magnetic fields assuming that only weak fields are present in the early universe. 
Those seed fields are then amplified by the astrophysical dynamo, including by galactic differential rotation,  
to the present $\mu$G level in the local ISM, 
comparable in energy density with thermal and turbulent energies.  
In present-day star formation, the magnetic fields indeed play colorful roles such as 
launching protostellar jets and outflows and transffering angular momentum in protostellar disks 
\citep[e.g.][]{machida08_jet}.
Recently, it has been recognized that, even in the primordial environment, 
magnetic fields may be amplified by so-called small-scale turbulent dynamo and 
the field strength may reach a dynamically significant level 
\citep{schleicher10,schober12,sur10,federrah11,turk12,latif13}.
Once the first stars are formed, magnetic fields can be generated and
amplified in the stars \citep[e.g.][]{spruit}, 
the circumstellar disks \citep{shiromoto} and even the neighboring prestellar cores \citep{ando,doi}. 
Finally, at their supernova explosions, 
those fields are dispersed into interstellar and intergalactic medium. 
Therefore, when the second-generation stars are formed from re-collapsed gas, 
magnetic fields are believed to be already present. 
Although proper assessment of the field strength in such environments is still elusive, 
it is worthwhile to study the effects of magnetic fields on low-metallicity star formation.

Recently, magneto-hydrodynamical (MHD) simulation of 
star formation in low-metallicity, magnetized clouds has been carried out
by \citet{peters14} 
under the {\it ideal} assumption, i.e., perfect flux freezing to the gas. 
They found that
the magnetic field generated via the small-scale dynamo stabilizes the
protostellar disk against fragmentation, 
counteracting destabilizing effect due to the enhanced cooling by metals.
However, it is still unclear in what metallicity range the ideal MHD approximation is 
justified. 
For a metal-free gas, it is known that the coupling between the field and gas is so tight that 
ideal MHD is a good assumption for the Jeans scale \citep{maki04,maki07}.
On the other hand, in the present-day solar metallicity gas, 
this approximation breaks down and the fields dissipate in the density range 
$10^{12}-10^{17}{\rm cm^{-3}}$.
As a result, the actual protostellar disk formation process can be different 
from that with the ideal approximation, which overestimates magnetic effects on dynamics. 
In view of this, we here investigate the
coupling between the gas and magnetic field, or in other words, 
resistivity of the gas, at various metallicities between 0 and $1Z_\odot$. 
Since it is the abundance of charged particles that controls the resistivity, 
we solve non-equilibrium chemical reaction network for prestellar cores, 
which are assumed to be collapsing dynamically.
We then examine the validity of the flux-freezing condition during the prestellar collapse. 

This paper is organized as follows: 
in Section 2, we describe the thermal/chemical evolution of the prestellar cores, 
which is used as a background model for the analysis of the magnetic-flux dissipation 
described in Section 3. Finally, in Section 4, we summarize our results and discuss its 
implication on the star formation in low-metallicity environments.

\hspace{1cm}
\section{Thermal and chemical evolution of collapsing prestellar cores}
\subsection{Numerical modelling}
We here employ one-zone model developed by
\citet{omukai05}, in which a spherical cloud with its core radius of 
the Jeans scale $R_{\rm J}$
is assumed to collapse in a runway, self-similar fashion, 
with some modification in chemical reactions. 
The calculated quantities in this model are those at the center. 
The density $\rho$ of the cloud increases as 
\begin{equation}
\frac{d\rho}{dt}=\frac{\rho}{t_{\rm ff}}\sqrt{1-f},
\end{equation}
where the free-fall time
\begin{equation}
 t_{\rm ff} = \sqrt{\frac{3\pi}{32G\rho}},
\end{equation}
$G$ the gravitational constant, and 
$f$ the ratio of the pressure gradient force to gravity at
the cloud center, which is fitted approximately as a function of
the effective ratio of specific heat $\gamma \equiv d\ln p/d\ln \rho$ ($p$ denotes the gas pressure):
\begin{equation}
f=\left\{
\begin{array}{l}
0~~~~~~~~~~~~~~~~~~~~~~~~~~~~~~~~~~~~~~~~~~~~\gamma < 0.83\\
0.6 + 2.5(\gamma -1)-6.0(\gamma-1)^2~~~ ~~0.83 < \gamma < 1\\
1.0+0.2(\gamma -4/3)-2.9(\gamma -4/3)^2~~ \gamma >1. 
\end{array}
\right.
\end{equation}
The thermal evolution is followed by solving the energy equation for  
internal energy per unit mass $\epsilon$:
\begin{equation}
\frac{d\epsilon}{dt}=-p\frac{d}{dt}\left(\frac{1}{\rho}\right)
 -\Lambda_{\rm net} \label{eq_energy},
\end{equation}
where $\Lambda_{\rm net}$ is the net cooling rate of the gas. 
The equations above are supplemented with the equation of state 
\begin{equation}
p=(\gamma_{\rm ad} -1)\rho \epsilon,
\end{equation}
where $\gamma_{\rm ad}$ is the adiabatic index.
The temperature $T$ and energy density $\epsilon$ are related with 
\begin{equation}
\epsilon = \frac{1}{\gamma_{\rm ad} -1} \frac{k_{\rm B}T}{\mu m_{\rm p}}
\end{equation}
where $k_{\rm B}$ is the Boltzmann constant, $\mu$ the mean molecular weight, and $m_{\rm p}$ 
the proton mass.
For the net cooling rate $\Lambda_{\rm net}$, 
in addition to the original cooling/heating processes in \citet{omukai05}, which 
include the radiative cooling by [CII], [CI], [OI], H$_2$, HD, CO, OH, H$_2$O lines, 
and by continuum from the primordial gas and dust, 
and cooling and heating associated with chemical reactions ({ see appendix}), 
we also consider the heating due to ionization by 
cosmic rays and radioactivity. 
The ionization heating rate is taken from \citet{wolfire} with 
assumption that the heat deposition rate associated with ionization by 
the decay of radioactive elements (REs) is the same as by cosmic rays.
{ We do not include the heating due to the dissipation of
magnetic fields. It has little impact on the thermal
evolution since the thermal energy is always larger than the magnetic energy
on average under realistic circumstances \citep[e.g.][]{tomida13}.}

The chemical fractions of the coolants are followed by solving 
the non-equilibrium chemical network. 
We add 14 new species 
Li, Li$^+$, Li$^{2+}$, Li$^{3+}$, Li$^-$, LiH,
LiH$^+$, M, M$^+$, G, G$^+$, G$^-$, G$^{2+}$, G$^{2-}$, 
which can be important charge providers, to the model of \citet{omukai05}.
Here, M and G stand for the metallic elements and grains, respectively. 
The reactions related to Li and its ions/molecules are listed in
\citet{bovino} and included in the present calculations.
The initial number abundance of Li relative to hydrogen, $y_{\rm Li}=n_{\rm
Li}/n_{\rm H}$, is taken
as $4.8\times 10^{-10}$, the observed ISM value in Small
Magellanic Cloud, consistent also with the big bang nucleosynthesis
\citep{howk12, galli_palla13}.
M and M$^+$ represent atoms and ions, respectively, of the metallic elements Na, Mg, Al, Ca, Fe and Ni, 
which are possible major electron providers. 
Since they all have similar rate coefficients for the charge transfer reaction
with other ions as well as for radiative recombination with electrons, 
we ignore their difference and treated collectively by summing up their elemental abundances. 
The number fraction of M is $y_{\rm M}=1.68\times 10^{-7}$ \citep{umebayashi90} at the solar metallicity 
and is proportionally reduced in lower metallicity cases.
The coefficients of M and M$^+$ related reactions are taken from \citet{umebayashi90} 
and \citet{prasad_huntress}.
Reactions relevant to the dust grains are crucial to determine the ionization state. 
We consider five ionization states
(G, G$^+$, G$^-$, G$^{2+}$, G$^{2-}$) of the dust grains. 
We assume the same mass fraction and size distribution as in \citet{omukai05}.
Mass fraction of the dust is
$0.939\times 10^{-2}$ below the water-ice evaporation temperature \citep{pollack94}, and
the same size distribution is assumed as \citep{mathis77}: 
\begin{equation}
\frac{dn_{\rm gr}}{da} \propto \left\{
\begin{array}{l}
a^{-3.5}~~~5\times 10^{-3} {\rm \mu m} < a < 1 {\rm \mu m} \\
a^{-5.5}~~~~~~~ 1{\rm \mu m} < a < 5 {\rm \mu m}.
\end{array}
\right.\label{eq_size}
\end{equation}
The evaporation of each component of the dust takes place  
instantaneously at its vaporization temperature \citep{omukai05, pollack94}.
The collision rates between grain-charged particles
and grain-grain are calculated by eqs.(3.1)-(3.5) of \citet{drain_sutin}, 
averaged over the size distribution (eq. \ref{eq_size}).

Ionization by cosmic rays and radioactivity controls the ionization degree in the clouds.
The cosmic rays ionize atoms and molecules either directly or indirectly by high-energy photons 
emitted at the direct ionization (described as CRPHOT). 
The direct ionization rates by cosmic rays are same as those of
\citet{omukai12}, except for the rates of M and HCO, which are taken from \citet{umebayashi90}.
The CRPHOT rates of M and HCO are taken from UMIST2 \citep{UMIST2}.
The attenuation of cosmic rays is considered for ionization rate by cosmic rays $\zeta_{\rm CR}$:
\begin{equation}
\zeta_{\rm CR} = \eta \zeta_{\rm CR,0}\exp\left(-\frac{\rho R_{\rm
				     J}}{\lambda}\right)
\end{equation}
where $\zeta_{\rm CR,0}$ is the cosmic-ray ionization rate in the local ISM and 
$\lambda$ the attenuation length. We use $\zeta_{\rm CR,0}=1\times
10^{-17} {\rm s^{-1}}$ and $\lambda = 96~ {\rm g~ cm^{-2}}$ in
the present calculations.
In the above, a parameter $\eta$ is introduced to specify the amount of ionization sources.
The core radius is given by the Jeans length 
\begin{equation}
 R_{\rm J} = \sqrt{\frac{\pi k_{\rm B} T}{G \mu m_{\rm p} \rho}}. \label{eq_jeans}
\end{equation}

REs also ionize the gas by emitting gamma rays at their decay. 
We classify REs into two categories by the decay time 
and treat their contribution to ionization rate differently.
We assume the amount of long-lived REs, e.g. $^{40}$K, with decay time $t_{\rm decay} \ga 1 $Gyr 
proportional to the total metallicity since the long-lived REs
accumulate in the ISM with little decay as the chemical enrichment proceeds. 
By using the value in the solar neighborhood for the solar metallicity case \citep{umebayashi09},
its contribution to the ionization rate is given by 
\begin{equation}
\zeta_{\rm RE,long} = 1.4\times 10^{-22} {\rm s^{-1}} Z/Z_\odot. 
\end{equation}
The short-lived REs, e.g.$^{26}$Al, have larger effects than the long-lived REs
since the gamma-ray intensity is inversely proportional to the decay time. 
The short-lived REs decay in shorter timescale than the age of the universe at high redshift and 
hence do not accumulate in the ISM. 
Without detailed knowledge of their amount, we here simply assume that 
it is also proportional to the same 
ionization parameter $\eta$ introduced above:
\begin{equation}
\zeta_{\rm RE,short} = 7.6\times 10^{-19} {\rm
s^{-1}} \eta. 
\end{equation}

In summary, the total ionization rate by the cosmic rays and the REs can be written as
\begin{equation}
 \zeta = \zeta_{\rm CR} + \zeta_{\rm RE,short} + \zeta_{\rm RE,long}.
\end{equation}
The first two terms are proportional to the parameter $\eta$ and the last term
is to the metallicity.
In this paper, we consider four cases with $\eta = 0, 0.01, 1, 10$, which 
are denoted as models 1, 2, 3  and 4, respectively.
Model 1 ($\eta = 0$) is the case without the ionization either by cosmic rays
or short-lived REs. In particular, for $Z=0$, no ionization source is present in this model, 
which can be regarded as the primordial pristine environment. 
Model 2 ($\eta = 0.01$) has 100 times smaller ionization rate than the local value.
\citet{stacy07} estimated the cosmic-ray intensity is about this value  
in first galaxies by summing
up the contribution from population III supernovae. 
\citet{nakauchi14} also assessed $\zeta_{\rm CR,0}$ at $6\la z\la
15$, spreading over two orders of magnitude around $10^{-19}{\rm s}^{-1}$.
Model 3 ($\eta = 1$) has the cosmic-ray intensity the same as in our Galaxy 
and allows us direct comparison with previous studies on 
present-day star formation such as \citet{nakano86}.
In model 4 ($\eta = 10$), with supernova explosions ten times more frequent than in our Galaxy, 
the environment would be similar to starburst galaxies at $z\la 5$ with 
much intense star-forming activities \citep{lacki14}.

We calculate nine cases with different metallicities 
$Z/Z_\odot=0,10^{-7},10^{-6},10^{-5},10^{-4},10^{-3},0.01,0.1,1$ for each ionization model.
Numerical convergence of non-equilibrium chemical reactions occasionally becomes very slow at high densities.
To avoid such difficulty, we switch to the equilibrium Saha equations of the H and
He system for $n_{\rm H} > 10^{18}{\rm cm^{-3}}$ neglecting all the metal species.
In some cases, the convergence becomes extremely slow 
at lower densities and we are forced to switch to the equilibrium calculation at 
$10^{16}{\rm cm^{-3}}<n_{\rm H}<10^{18}{\rm cm^{-3}}$.
In metal-enriched cases, this treatment causes a discontinuity in the fraction 
of some charged species around $10^{18}{\rm cm^{-3}}$. 
{ Those discontinuities comes from the fact that we are still omitting 
some reactions required to achieve the chemical (i.e., Saha) equilibrium 
such as the ionization by thermal photons, etc. 
Note that to reach the correct Saha equilibrium at high enough density, both 
the forward and reverse reactions must be included for all the relevant processes. 
In addition, we do not include the species other than e$^-$, H, H$^+$,
He, He$^+$, He$^{++}$ in solving the Saha equations, which can be part 
of the reason of the discontinuity.}
\footnote{ Vaporization of the grains also generate
discontinuities at $\sim 10^{16}{\rm cm^{-3}}$, because we model the
vaporization of grain materials by the threshold temperatures. Hence the
discontinuities around $\sim 10^{16}{\rm cm^{-3}}$ is reasonable.}

However, this does not affect significantly our conclusion on the density range of 
magnetic field dissipation. This is because the discontinuity appears at high
enough density where the charged dust is already vaporized. In the
absence of charged grains, electrons dominate the electric current
carrier, which is more or less properly estimated by the Saha equation
in such a high density regime.

\subsection{Thermal evolution}
\begin{figure*}
\plotone{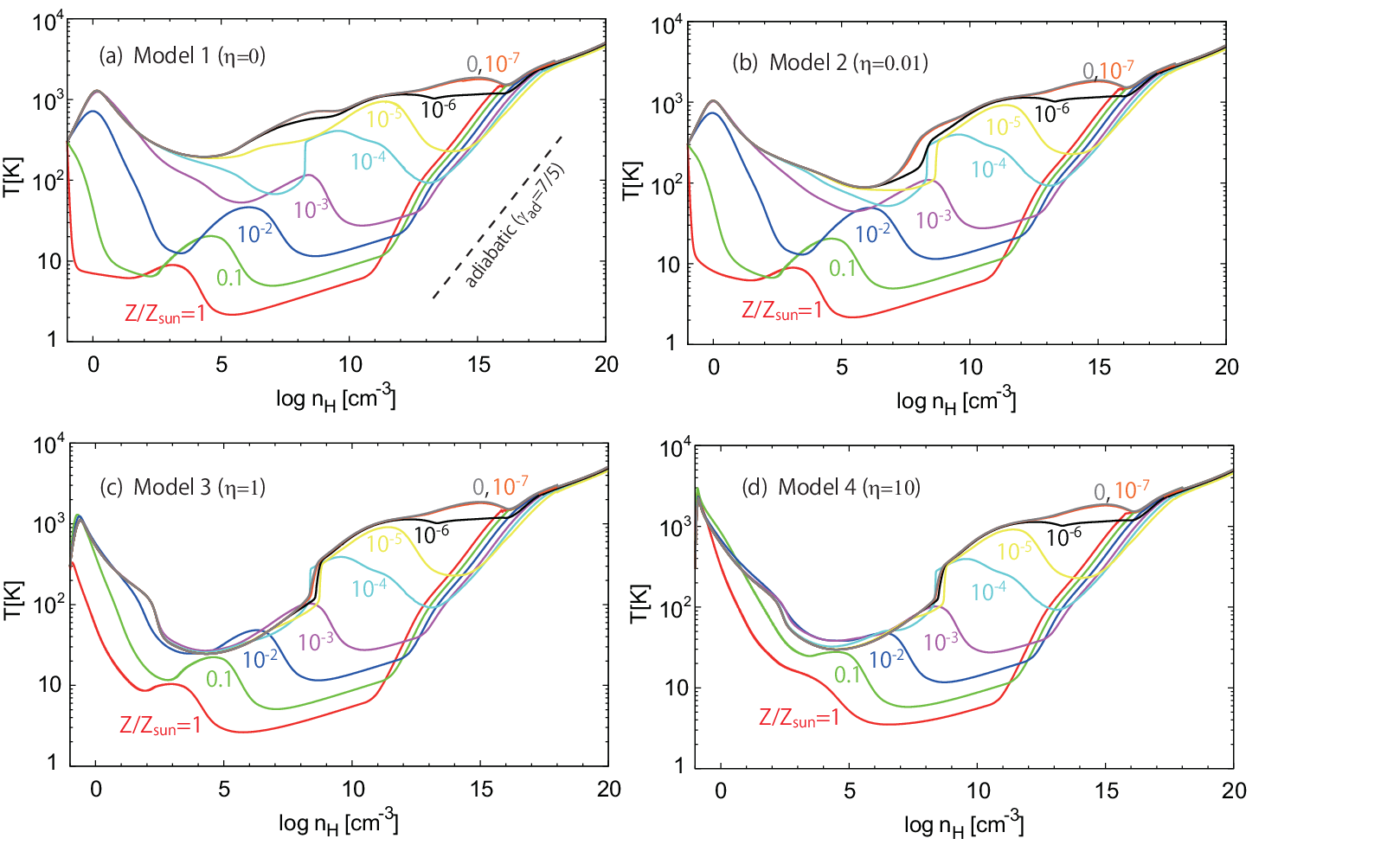}
\caption{Temperature evolution of the prestellar cores 
for different metallicities as a function of the number density. 
Four panels (a)-(d) correspond to cases with four different ionization parameters 
$\eta =0, 0.01, 1$ and $10$, respectively. The values of metallicity are indicated by numbers in the Figure.
In panel (a), the adiabatic temperature gradient for $\gamma_{\rm ad}=7/5$ is shown by the dashed line.
}
\label{fig_nT}
\end{figure*}
In this section, we describe the thermal evolution of the clouds for the four ionization models. 
The panels in Fig.\ref{fig_nT} show the evolutionary tracks on the $n_{\rm H}-T$ plane 
for models 1 - 4 ($\eta=0, 0.01, 1$ and 10).
Different curves in each panel are for different metallicities.
Below $n_{\rm H}\la 10^{15}{\rm cm}^{-3}$, the temperature decreases with increasing metallicity 
owing to the larger amount of coolants for a given ionization parameter.
After the central part of the cloud becomes optically thick to the dust thermal radiation,
all the evolutionary tracks with different metallicities converge to a single track 
for $\ga 10^{16}{\rm cm^{-3}}$ \citep{omukai00}.

With larger ionization parameter, the temperature tends to be lower for the same 
metallicity and density range.
For example, with a small amount of cosmic-ray ionization (model 2, $\eta=0.01$), the temperature is lower than 
in the no-ionization case (model 1, $\eta=0$) in the range $10^5-10^9{\rm cm^{-3}}$. 
This tendency is more clear in models with higher ionization rate, i.e., 
models 3 ($\eta =1$) and 4 ($\eta=10$). 
This is because ionization promotes H$_2$ formation via the electron catalyzed reaction. 
By enhanced H$_2$ cooling, the gas reaches to lower temperature, which is favorable for HD formation.   
Hence, the temperature plummets further by the HD cooling with
its excitation energy four times lower than H$_2$ \citep{stacy07,nakauchi14}.
With $Z\ga 10^{-3}Z_\odot$, other cooling agents such as C and O are more important 
than HD and the temperature evolution is little affected by the difference in ionization rate. 
In all cases, the temperature in high-density regime ($n_{\rm H} \ga 10^{10}{\rm cm^{-3}}$) 
does not depend on the ionization parameter
because of cosmic-ray attenuation as well as dominance of dust cooling.

\subsection{Chemical abundances}
\begin{figure*}
\plotone{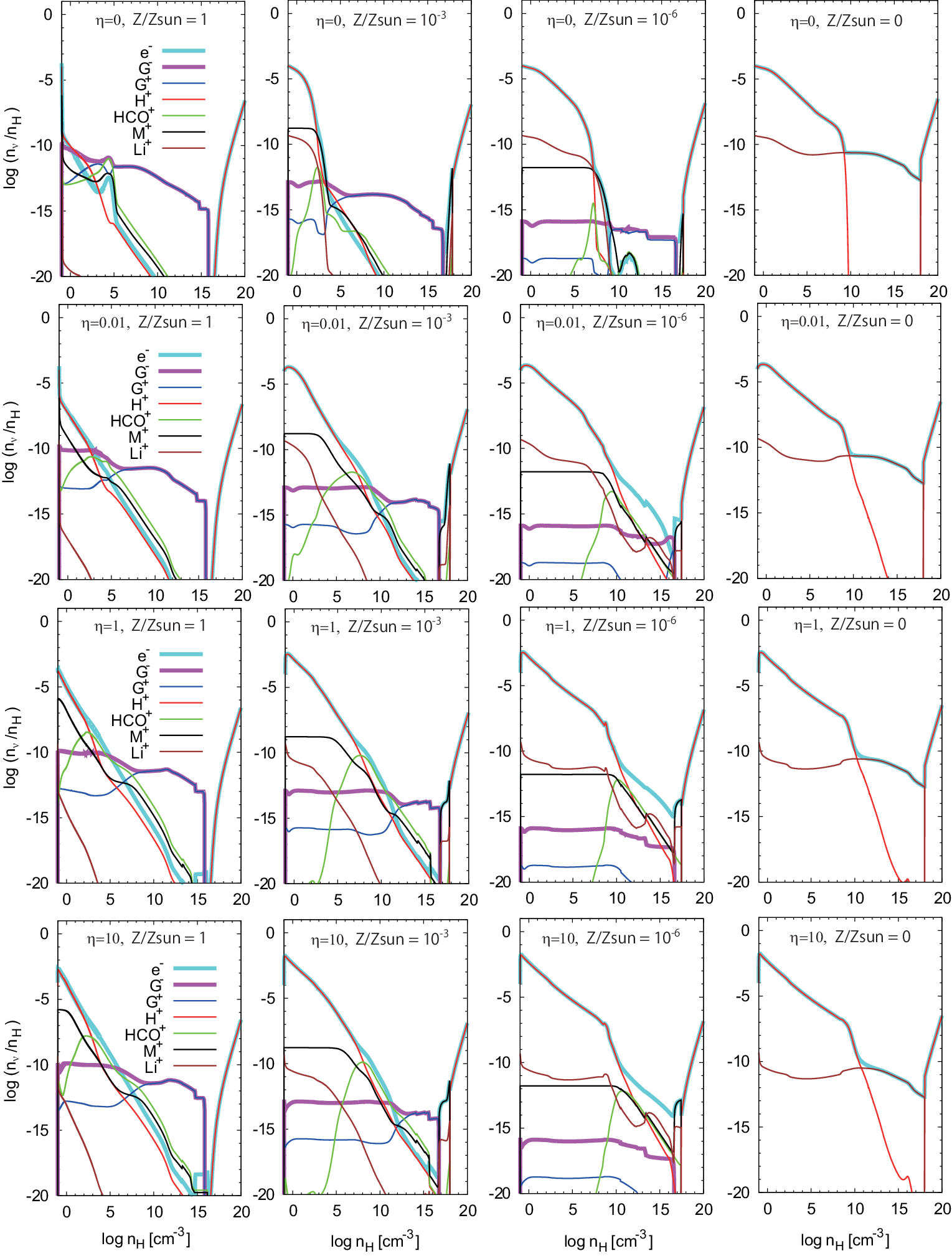}
\caption{Abundance of major charged species as a function of
 number density.  Results for ionization parameter $\eta=0, 0.01, 1$ and $10$ (models 1 - 4) 
are shown from the top to the bottom rows, and for metallicities $Z=Z_\odot, 10^{-3}Z_\odot,
 10^{-6}Z_\odot, 0$ from the left to the right columns.
The negatively charged species e$^{-}$ and G$^-$ are shown by thick solid curves,
while the positive ions (H$^+$, G$^+$, HCO$^+$, M$^+$, Li$^+$) are plotted by thin solid lines.
Note that the discontinuities seen at $\sim 10^{16}{\rm cm^{-3}}$ for metal-enriched cases are 
artifacts caused by our switching from non-equilibrium to equilibrium chemistry. 
}
\label{fig_ny}
\end{figure*}

Since we are interested in the resistivity of the gas, we plot 
the abundances of 7 species (e$^-$, G$^\pm$, H$^+$, HCO$^+$, M$^+$, Li$^+$), 
which can be important charge carriers, as a function of number density in Fig. \ref{fig_ny}.
In this Figure, 16 panels correspond to models with 4 different
metallicities ($Z/Z_{\odot}=1, 10^{-3}, 10^{-6}$ and 0) and 4 ionization rates 
($\eta=0, 0.01, 1$, and 10). 
In all cases, e$^-$ and H$^+$ are the dominant charged particles in the low-density regime, 
but their recombination proceeds with increasing density. 
The charged grains G$^\pm$ eventually become the dominant charged component 
at some density except in the zero-metallicity cases (i.e., the rightmost column).
At even higher densities ($\ga 10^{16}{\rm cm^{-3}}$), the vaporization of grains
quenches the recombination on the dust surface, and thus the ions and electrons dominate the charge again. 
The behavior in the zero-metallicity cases is quite different.
The initial decrease of the electron fraction by the recombination with 
H$^+$ is followed by a floor set by Li$^+$. 
Hence, the ionization degree never becomes lower than $10^{-12}$ for all $\eta$ values.
Note that this behavior of the ionization degree in zero-metallicity clouds has been reported 
for smaller ranges of ionization parameter \citep{maki04,maki07,schleicher10}. 

Next we see the difference among the ionization models. 
In the $\eta=0$ case (model 1; first row), the recombination proceeds much faster than in non-zero $\eta$ cases 
(the second to the fourth rows). 
The electron fraction drops exponentially as a function of the density for the $\eta=0$ cases (model 1), 
while it decreases more slowly in other cases as a result of the cosmic-ray/RE ionization.
At a fixed density, the electron fraction increases with the ionization rate. 
The higher electron fraction means the less importance of charged grains:   
the density range where the grains dominate the charge shrinks toward increasing $\eta$. 
For cases with $\eta = 1$ and $10$ (models 3 and 4) and with $10^{-6}Z_\odot$, the charged grains 
never dominate the charge because of the small amount of dust and relatively high ionization rate. 

\section{Dissipation of magnetic flux}
Dissipation of magnetic fields in a gas can be studied 
by estimating the ambipolar diffusion rate and Ohmic resistivity. 
Here we employ the formalism developed by \citet{nakano86},
which allows us to handle the two dissipation processes simultaneously. 

The drift velocity of magnetic field lines relative to the gas is given by \citep{nakano86}:
\begin{equation}
 v_{{\rm B}x}=\frac{A_1}{A}\frac{1}{c}\left(\bm{j}\times\bm{B}\right)_x,\label{eq_vB}
\end{equation}
where
\begin{eqnarray}
 A&=&A_1^2+A_2^2,\\
A_1&=& \sum_\nu \frac{\rho_\nu\tau_\nu\omega_\nu^2}{1+\tau_\nu^2\omega_\nu^2},\\
A_2&=& \sum_\nu \frac{\rho_\nu\omega_\nu}{1+\tau_\nu^2\omega_\nu^2}.
\end{eqnarray}
Here $c$ is the speed of light, $\bm{j}$ and $\bm{B}$ the
electric current and magnetic flux density, 
$\rho_\nu$ and $\omega_\nu$ are the density and the cyclotron frequency
of the charged particle $\nu$, respectively.
The subscript $x$ denotes the direction parallel to $\bm{j}\times\bm{B}$.
The viscous damping timescale of the relative
velocity between charged ($\nu$) and neutral (n) particles is given by
\begin{equation}
\tau_\nu =\frac{\rho_\nu} { \mu_{\nu {\rm n}}  n_\nu n_{\rm n} 
\bra\sigma v\ket_{\nu {\rm n}} }
\end{equation}
where $\mu_{\nu {\rm n}}$, $n_\nu$, $n_{\rm n}$ are the reduced mass
of particles $\nu$ and ${\rm n}$, number densities of $\nu$ and ${\rm n}$ species, 
respectively. 
The momentum-transfer rate coefficient via the collision between
$\nu$ and n, $\bra\sigma v\ket_{\nu {\rm n}}$, 
is evaluated as in \citet{pinto_galli} and averaged over the dust size distribution (\ref{eq_size}).
Here, we replace the term $\left(\bm{j}\times\bm{B}\right)_x$ in
equation (\ref{eq_vB}) by the mean value of magnetic force $B^2/4\pi R$, where $B$ 
is the mean magnetic field strength and $R$ the scale length that we are interested 
in \citep{nakano86}, which is taken as the Jeans radius $R_{\rm J}$ (eq. \ref{eq_jeans}). 

Relative importance of the Ohmic loss to the ambipolar diffusion is
determined by the quantity $|\tau_\nu \omega_\nu|$. 
If it is less than unity, the Ohmic loss is more important than the ambipolar diffusion, 
and vice versa. 
Approximate expressions for two limits can be given 
as \citep{nakano86}:
\begin{equation}
 v_{{\rm B}x}\sim \left\{
\begin{array}{l}
\displaystyle \frac{c^2}{4\pi \sigma_{\rm c}R}~~~~(|\tau_\nu
 \omega_\nu|\ll 1 ~~~{\rm Ohmic~loss})\label{eq_vb_ohmic}\\
\displaystyle \frac{\tau_{\rm i}}{\rho_{\rm i}}\frac{B^2}{4\pi R}
 ~~~~(|\tau_\nu \omega_\nu|\gg 1~~~{\rm ambipolar~diffusion})\label{eq_vb_AD}
\label{eq:vBx}
\end{array}
\right.
\end{equation} 
where $\sigma_{\rm c}$ is the conductivity defined as
\begin{equation}
 \sigma_{\rm c} =\sum_\nu \frac{q_\nu\tau_\nu n_\nu}{m_\nu},
\end{equation}
and $q_\nu$ and $m_\nu$ are the charge and the mass of a charged particle $\nu$, respectively. 
The subscript i represents the dominant ion species.

We compare the drift velocity $v_{{\rm B} x}$ (eq. \ref{eq_vb_AD}) 
with the free-fall velocity $u_{\rm ff}$
defined as
\begin{equation}
 u_{\rm ff} \equiv \sqrt{\frac{2GM}{R}}. \label{eq_uff}
\end{equation}
For the mass and radius of the prestellar core $M$ and $R$, we use 
the instantaneous Jeans mass ($M_{\rm J}=4\pi\rho R_{\rm J}^3/3 $) and
radius (eq.\ref{eq_jeans}), respectively. 
If $v_{{\rm B}x} > u_{\rm ff}$, the magnetic field lines move away 
from the core by their own tension before the significant collapse, i.e., the magnetic fields dissipate.
On the other hand, for $v_{\rm Bx} < u_{\rm ff}$, 
the magnetic fields hardly dissipate and the field lines can be regarded frozen to the cloud.

\begin{figure*}
\plotone{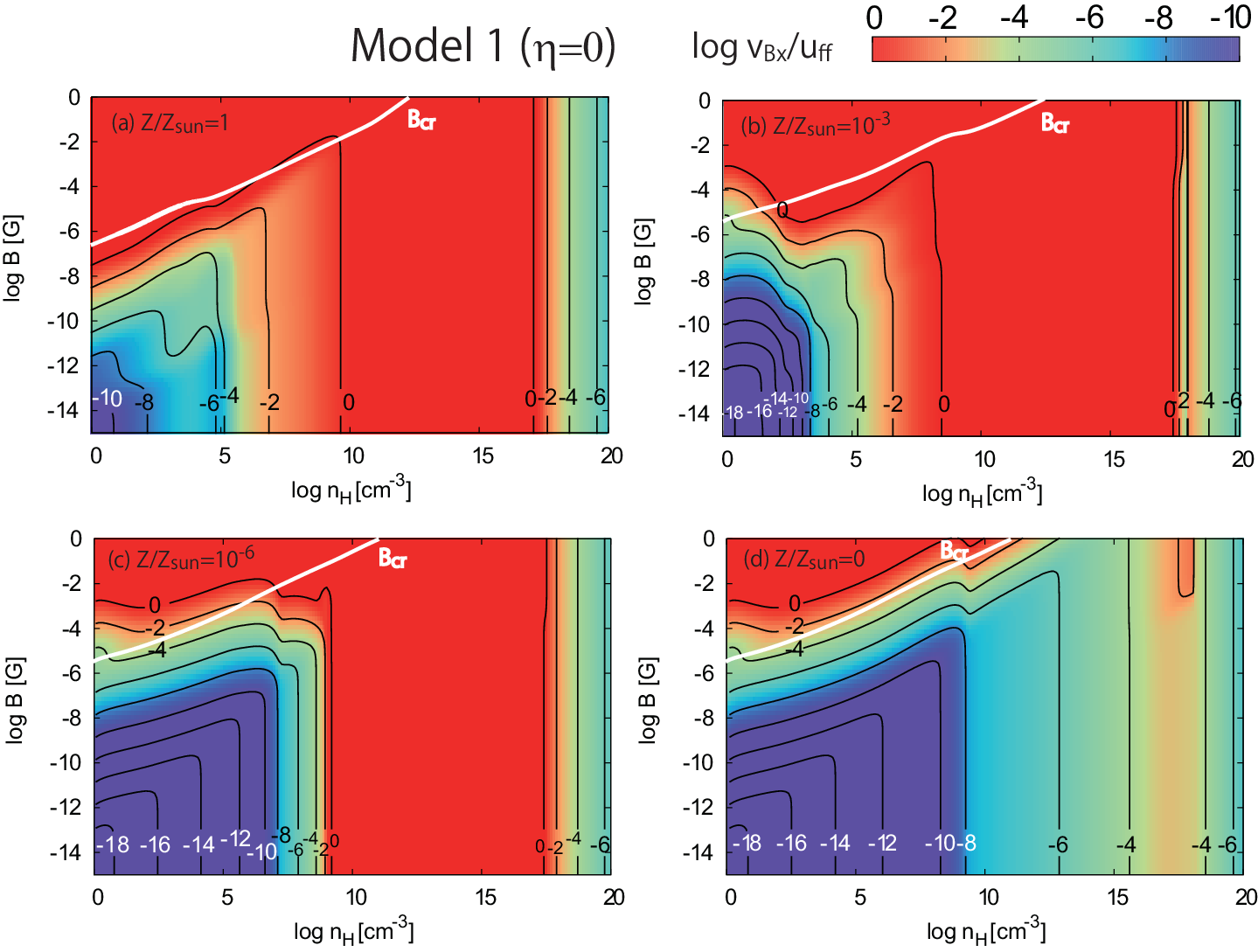}
\caption{The ratio $v_{{\rm B}x}/u_{\rm ff}$ on the $n_{\rm H}-B$ plane for the $\eta=0$ cases (model 1). 
The numbers indicated on the contours denote its logarithmic values. Four panels (a)-(d)
 correspond to different metallicities (a)$Z/Z_\odot = 1$, (b)$10^{-3}$,
 (c)$10^{-6}$, and (d)$0$, respectively. The white curves denote $B_{\rm
 cr}$(see text).}
\label{fig_cont_model1}
\end{figure*}
\begin{figure*}
\plotone{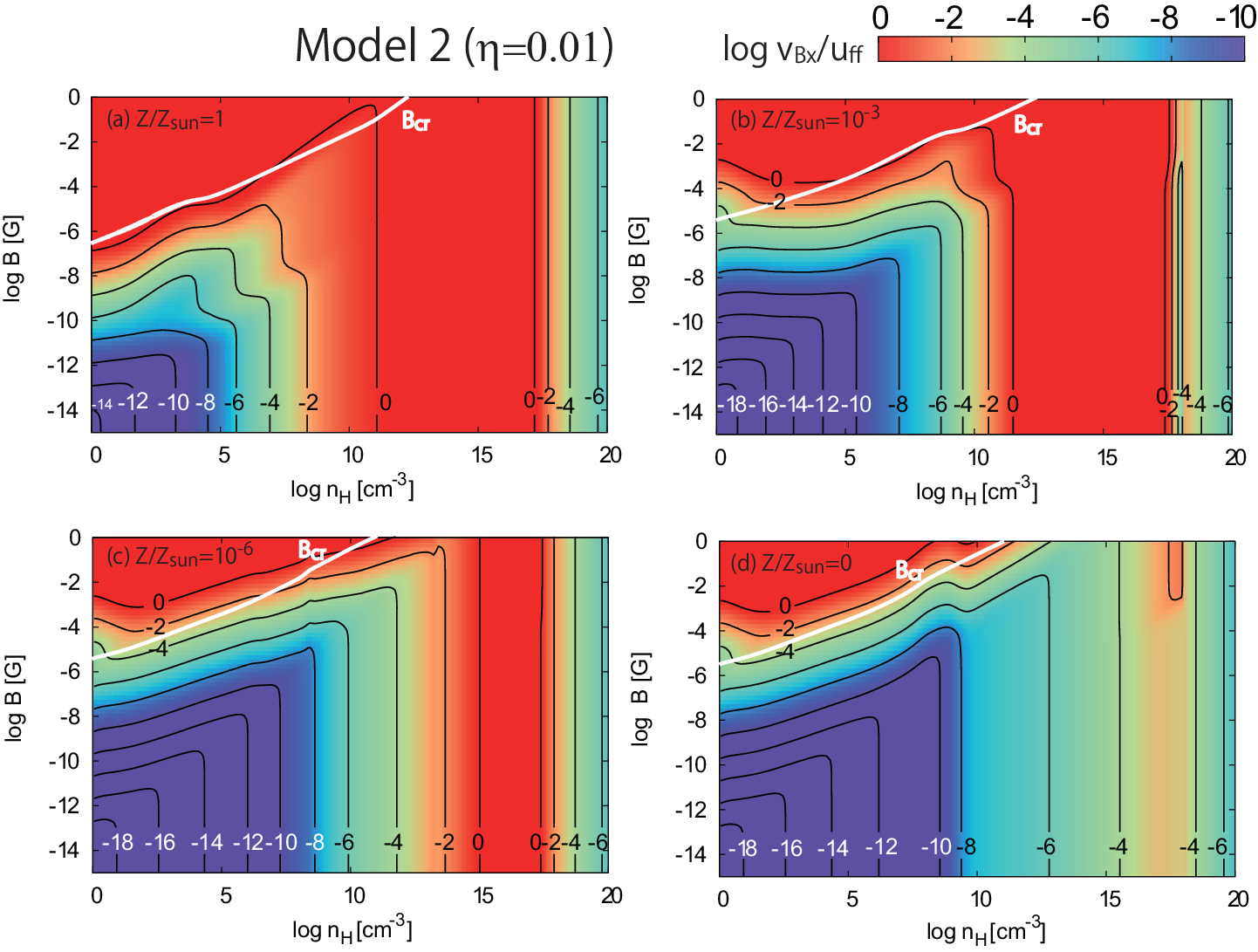}
\caption{Same as Fig.\ref{fig_cont_model1}, but for
the ionization parameter $\eta =0.01$ (model 2).}
\label{fig_cont_model2}
\end{figure*}
\begin{figure*}
\plotone{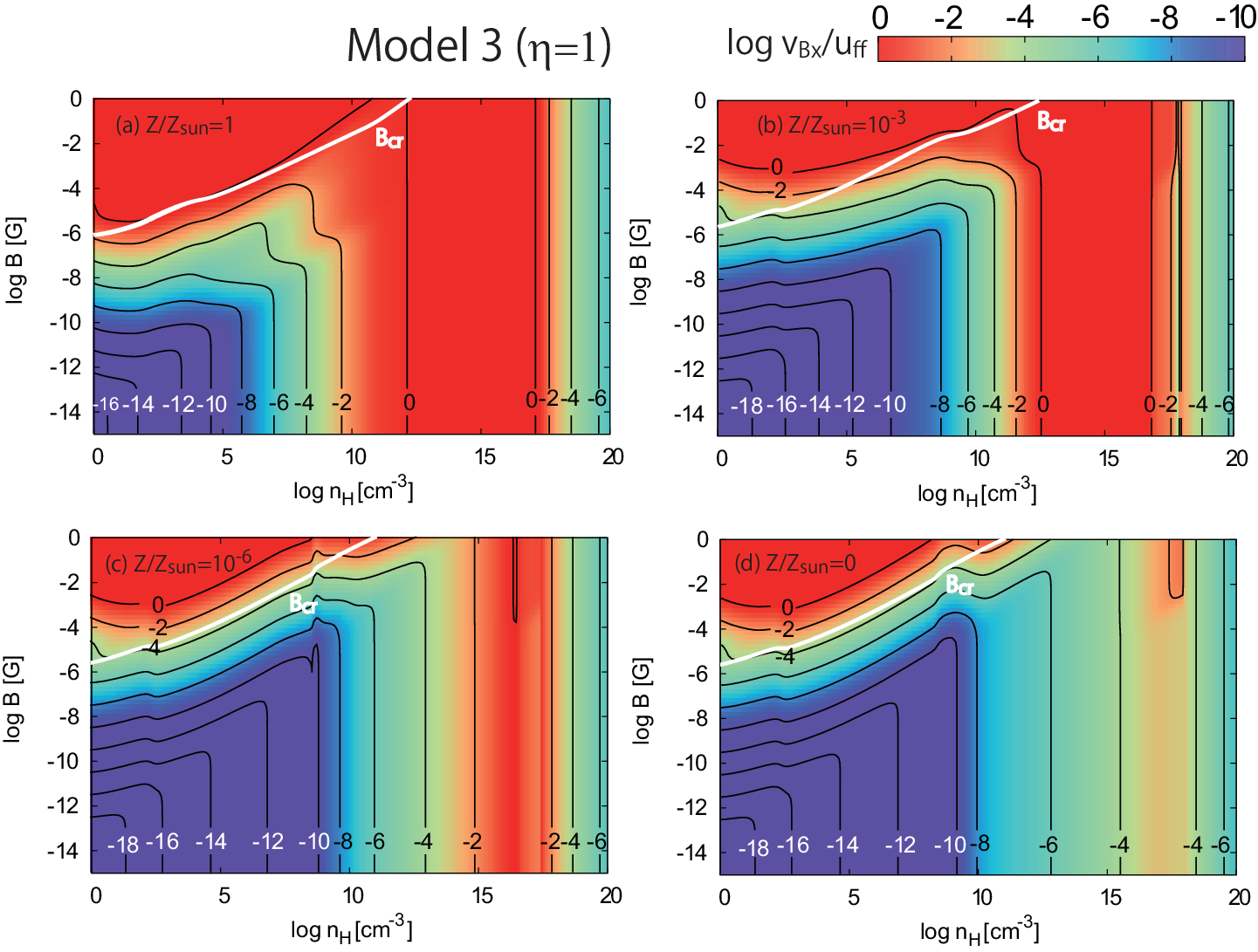}
\caption{Same as Fig.\ref{fig_cont_model1}, but for
the ionization parameter $\eta=1$ (model 3).}
\label{fig_cont_model3}
\end{figure*}
\begin{figure*}
\plotone{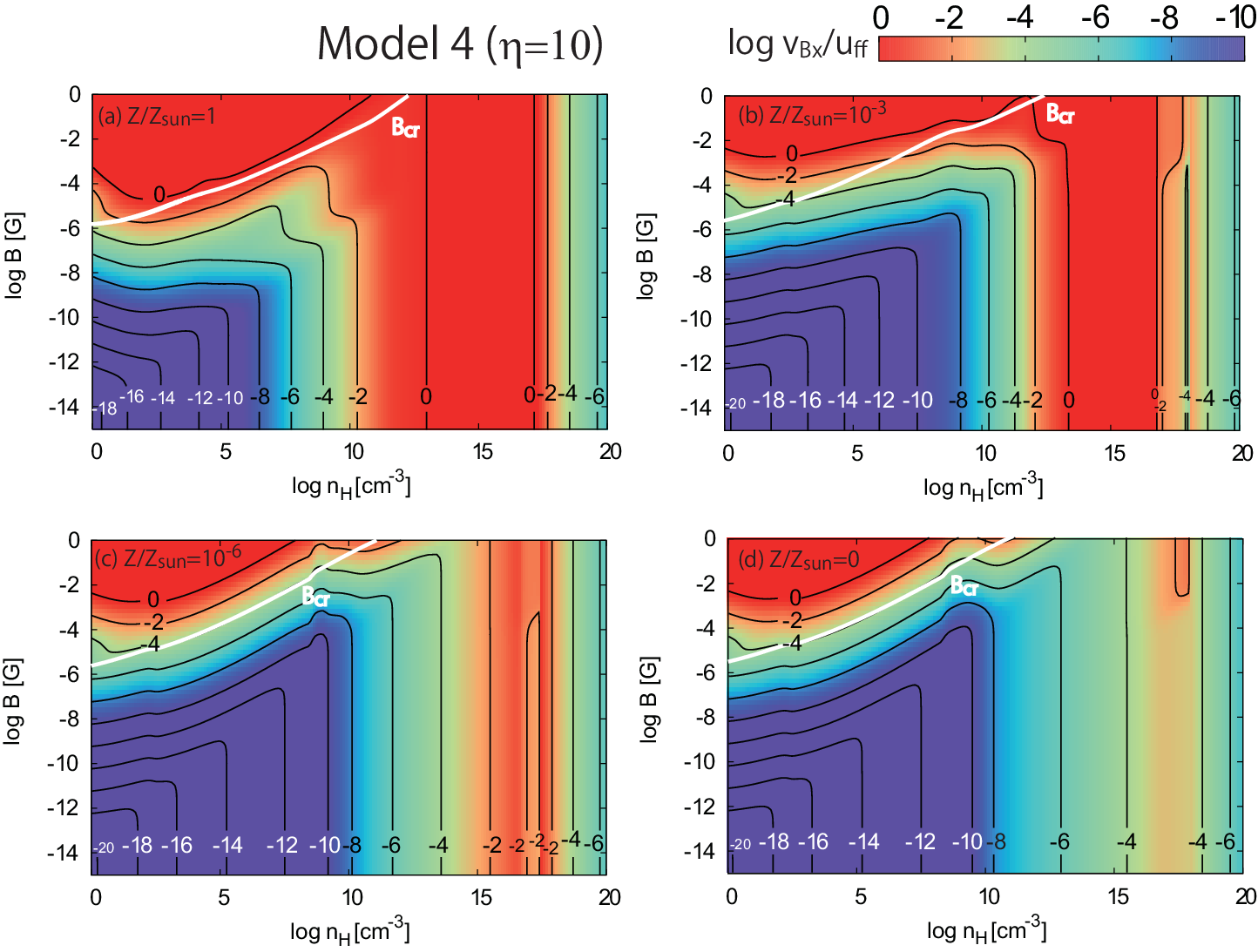}
\caption{Same as Fig.\ref{fig_cont_model1}, but for
the ionization parameter $\eta=10$ (model 4).}
\label{fig_cont_model4}
\end{figure*}

Figs.\ref{fig_cont_model1}-\ref{fig_cont_model4} show the ratio 
$v_{{\rm B}x}/u_{\rm ff}$ on the $n_{\rm H}$ -$B$ plane for four cases
with different ionization parameters 
$\eta =0, 0.01, 1$ and 10 (i.e., models 1-4, respectively). 
In each figure, four panels show the cases with metallicities
$1Z_\odot$, $10^{-3}Z_\odot$, $10^{-6}Z_\odot$ and $0$.
In the red regions in those panels,  the condition $v_{{\rm B}x}/u_{\rm ff} \ga 1$ is satisfied,
i.e. the magnetic flux dissipates from the gas during the collapse. 

We describe the results for the $\eta=1$ case (Fig.\ref{fig_cont_model3}: model 3) 
as the fiducial model in the followings. 
The qualitative behavior is common in all the models although with some quantitative differences.
First, in relatively metal-enriched cases $Z_\odot$ and $10^{-3}Z_\odot$, i.e., top two panels (a) and (b), 
a red strip parallel to the $y$-axis appears in the density range $10^{12}-10^{17}{\rm cm}^{-3}$, i.e.,  
the magnetic fields dissipates between this interval irrespective of the field strength $B$.
This is due to the Ohmic loss, for which the drift velocity $v_{\rm Bx}$ does not depend on $B$ 
(see eq. \ref{eq_vb_ohmic}), and so the dissipation occurs regardless of the field strength.
This strong dissipation comes from the dominance of the grains in charge 
in this density range as seen in Fig. \ref{fig_ny}.
The grains have much larger inertia than the electrons, causing
large resistivity to the current. 
Existence of such a dissipation density range has already been pointed out
by \citet{nakano86} for the local ISM condition. 
Our study reveals that it continues to exist even for lower metallicity gases.
In extremely metal-poor cases (see panels c and d ) for $Z=10^{-6}Z_\odot$ and 0, respectively), 
the dissipation range becomes very narrow for $Z=10^{-6}Z_\odot$ (panel c) and almost disappears finally 
for $Z=0$ (panel d) because the grains never dominate the charge 
owing to their small amount (see Fig.\ref{fig_ny}) and so the Ohmic dissipation does not work. 
In addition, the recombination of charges on the grain surface does not proceed efficiently 
at such low metallicity, and the abundance of electrons and ions remains relatively high. 
In particular, in the $Z=0$ case (panel d), Li$^+$ remains in the gas phase because of
the complete absence of dust grains, which would absorb Li$^+$ otherwise. 
As a result, the electron supply from Li$^+$ gives a floor to the electron fraction, 
which keeps the resistivity very low and diminishes the dissipation range.

In all the panels in Figs.\ref{fig_cont_model1} - \ref{fig_cont_model4}, 
dissipation regions by the ambipolar diffusion are present at 
the upper left (low density and strong field) corners. 
From equations (\ref{eq_vb_AD}) and (\ref{eq_uff}) at the Jeans scale, 
the boundary of the dissipation region defined as $v_{\rm Bx}/u_{\rm ff} = 1$ is given by 
\begin{equation}
B\propto n_{\rm H}^{3/4} T^{1/2},
\end{equation} 
roughly consistent with the slope of the contour. 
Note that the magnetic energy density is comparable to the gravitational energy density, 
and the collapse is prohibited in this parameter range.
The critical magnetic flux $B_{\rm cr}$, which is defined by the relation
\begin{equation}
B_{\rm cr}^2/4\pi R_{\rm J} =  \rho GM_{\rm J}/R_{\rm J}^2,
\end{equation}
is superimposed at the upper left corner of each panel by a white curve.

This means that physical condition in this region is not suitable for star-forming clouds 
and the ambipolar diffusion is not important in most cases.
For some cases with high metallicity ($\ga 10^{-3}Z_{\odot}$) and low ionization parameter
($\eta=0$ or $0.01$), 
this boundary shifts toward smaller $B$ into a region where 
the cloud $v_{{\rm B}x}/u_{\rm ff} = 1$ is able to collapse. 
In such cases, the ambipolar diffusion can be important. 

\begin{figure}
\plotone{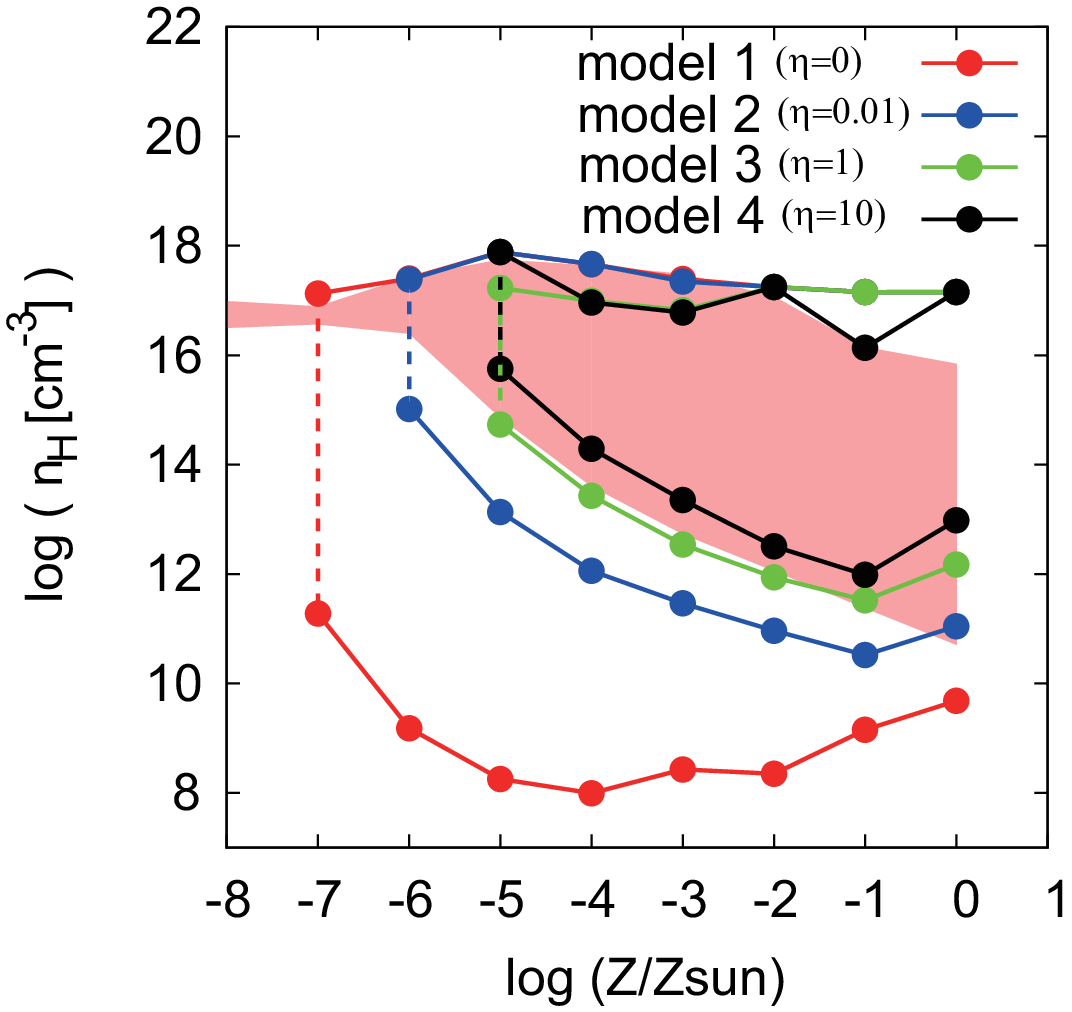}
\caption{Density range where the magnetic fields dissipate by the Ohmic
 loss for the gas with different metallicities.
Two curves of the same color indicate the lower/upper bound of the range.
Four colors correspond to different ionization parameters $\eta=0$,
 0.01, 1, and 10. Vertical dashed lines represent the lower bound of
 the metallicities below which no dissipation regions are found.
Also shown for comparison is the density range for the first-core phase (shaded region), 
where the protostellar outflow is expected to be launched if significant magnetic fields are present.  
Note that the first-core density range does not depend on the ionization parameter. 
 }
\label{fig_cZ}
\end{figure}

For the Ohmic dissipation, whose rate depends only on the density,  
we can define dissipation density range where $v_{\rm Bx}/u_{\rm ff} > 1$.  
Fig. \ref{fig_cZ} shows the upper/lower bound of the dissipation density range by the Ohmic loss 
as a function of metallicity for different ionization parameters (indicated by 
lines with different colors).
Except for the case of $\eta = 0$, the gas becomes dissipative from 
higher density, i.e., the lower bound of the dissipation range increases,
with decreasing metallicity for $Z < 0.1 Z_{\odot}$. 
This is because, for lower metallicity and smaller amount of the dust, 
the grains dominate the charge only at higher density (see Fig. \ref{fig_ny}). 
On the contrary, without ionization source $\eta=0$, the lower bound 
increases with metallicity for $\ga 10^{-2}Z_{\odot}$. 
This is because the charge is carried by the dust grains in this case
even at low densities ($n_{\rm H}\sim 10^{5}{\rm cm}^{-3}$, see
Fig. \ref{fig_ny}) due to the rapid drop-off of the electron fraction. 
As a result, the larger amount of charge carriers is available for  
the larger metallicity and so the gas becomes less resistive. 
The same effect is also responsible for the upturn of the lower boundary around 
the solar metallicity in the cases with $\eta >0$.
On the other hand, the upper bound of the dissipation range 
remains almost the same for all models. 
Consequently, the dissipative range becomes narrower with decreasing metallicity.
Eventually, in the case with $\eta=1$ ($\eta=10$), 
for $<10^{-7}Z_\odot$ ($< 10^{-6}Z_\odot$, respectively) 
the dissipation range disappears. 
Such clouds do not experience the dissipative phase during the collapse.
In addition, we can see that the dissipative phase is shorter 
in the density range for the cases with larger ionization parameter
because of the larger amount of charge carriers. 

\section{Summary and Discussion}
\label{discussion}
We have studied the coupling between the magnetic fields and the gas in the range of metallicity $Z=0-1 Z_{\odot}$ and ionization rate
by cosmic rays/radioactive elements $0-10$ times the solar neighborhood value.
For this purpose, we have calculated the thermal and chemical evolution of
prestellar cloud cores with the Jeans scale by solving non-equilibrium chemical reactions 
for important charge carriers including charged grains.
We have found that, for the Milky-Way like environment with metallicity $Z_{\odot}$ and 
the cosmic ionization rate $10^{-17}{\rm s^{-1}}$, the magnetic flux at the scale of the core
dissipates from the gas by the Ohmic loss, regardless of the field strength, 
in the density range $10^{12}{\rm cm^{-3}} \la n_{\rm H} \la 10^{17}{\rm cm^{-3}}$.
This dissipation is due to the enhanced resistivity in this range by the dominance of grains as 
charge carriers. 
With decreasing metallicity and increasing ionization rate, more charge is carried by electrons and 
ions relative to grains and, as a result, the dissipation density range by the Ohmic loss becomes narrower. 
For metallicity less than $10^{-6}-10^{-7}Z_\odot$, depending on the ionization rate, 
the magnetic flux is always frozen to the gas at the Jeans scale. 
Hence, the magnetic field present in the prestellar core with such low metallicity  
 will be taken into the forming protostar without dissipation.  

In what follows, we discuss implications of the magnetic dissipation 
for the dynamics of star-forming clouds, as well as  
for the small-scale dynamo action in the metal-free case. 
\\

\subsection{Implications for the dynamics of prestellar cores}

In the Milky-Way like environment, temperature in the prestellar core 
increases nearly adiabatically in the range $10^{11}{\rm cm^{-3}} \la n_{\rm H} \la 10^{16}{\rm
cm^{-3}}$ (see Fig. \ref{fig_nT}c).
In this period, increasing pressure stops the dynamical collapse and 
a hydrostatic core, which is called the first core, 
forms at the center. The first core subsequently contracts only quasi-statically 
with accretion of matter from the envelope.
Because of the slow radial velocity, the first core rotates many times before significant contraction.  
If the magnetic field is present and tightly coupled to the gas in this phase, 
the field lines are twisted enormously. 
The enhanced magnetic pressure/tension force launch the outflows from the first core, 
as well as accelerates the cloud collapse by transffering angular momentum efficiently
by the magnetic breaking. 
On the other hand, if magnetic fields dissipate before the first-core phase, 
neither outflows nor the magnetic breaking will be operative \citep{machida08_jet}.

In Fig. \ref{fig_cZ}, the density ranges of the first core are indicated along with 
the Ohmic dissipation range for comparison. 
The shaded region represents the range of adiabatic phase 
$\gamma \ge \gamma_{\rm ad}$, where the first core is formed. 
Note that the density range of the first core is independent of the
ionization models.
In the case of no ionization source ($\eta=0$), the dissipation zone extends
in the density range much wider than that of the first core for all metallicities. 
Hence, MHD effects such as the outflow launching or the magnetic breaking 
would not work in the first-core phase.  
Note that the first-core phase disappears at lowest metallicities, $\la 10^{-6}Z_\odot$.  
The upper density bound of the first-core phase marks the onset of another 
dynamical collapse, which is induced by the effective cooling by the H$_2$ dissociation. 
Eventually, for density $n_{\rm H}\ga 10^{20}{\rm cm}^{-3}$,
following the completion of H$_2$ dissociation,  
the temperature begins increasing adiabatically again. 
The collapse slows down and another hydrostatic core, called the second core or protostar, forms.
In this phase, the magnetic coupling is tight and so the MHD effects can be important,  
as long as some magnetic fields are still present, for all metallicities and ionization rates. 
With $\eta=0.01$ and $Z\la 0.1 Z_\odot$, 
i.e., in the first-galaxy-like environment \citep[e.g.][]{wise12},
the magnetic dissipation still begins before the first core formation.
This means that the MHD effect in such an environment 
would be weaker than in the present-day cases.
On the other hand, with ionization rate similar to the Milky-Way value ($\eta=1$),
the magnetic dissipation begins almost at the same density as 
the onset of the first core phase $Z<0.1Z_{\odot}$
or even later than that for $Z>0.1Z_{\odot}$. 
In the latter case, the magnetic coupling is still tight in 
the early first-core phase, and thus rotation of the first core twists 
the field lines at its edge. 
Hence, above $0.1Z_{\odot}$, the magnetic effects are important on the dynamics
in the first-core phase as is known from studies on the present-day star formation.
Finally, with ionization rate higher than the local value ($\eta=10$, 
starburst galaxy-like environment), the dissipation zone becomes even narrower. 
The dissipation begins always after the first-core formation 
at any metallicity, and the dissipation zone
eventually disappears for $Z \la 10^{-6}Z_\odot$.
Hence the MHD effect is always important as long as the first core forms.

We plan to carry out numerical MHD simulations of low-metallicity star formation 
by utilizing the resistivity obtained in this calculation as a future study.

\subsection{Implications for small-scale dynamo action in primordial gas}

Recently, amplification of magnetic fields by the small-scale dynamo 
during the first-star formation has attracted attention of some authors 
\citep[e.g.][]{schober12}. 
This theory assumes that parental clouds of the first stars are highly turbulent, 
as in the present-day molecular clouds, down to the viscous scale much smaller than the Jeans length. 
In such a circumstance, the turbulent energy will be transferred
to magnetic energy within the eddy timescale of the turbulence,
if the magnetic field is tightly coupled to the gas \citep{kazantsev,brandenburg05}. 
A magnetic field at the smaller scale has shorter eddy timescale and 
is amplified faster than that at the larger scale.
The energy of magnetic field at the viscous scale is thus amplified most quickly. 
The amplified magnetic field at a smaller scale is supposed to inversely cascades to larger scales. 
Finally, this continues until the magnetic field reaches the equi-partition level even 
at the Jeans scale.

Now, we examine the validity of the flux freezing in
the primordial gas, which is the basic assumption of the small-scale dynamo theory. 
As shown in Fig. \ref{fig_cont_model1}d, 
for $\eta=0$ (i.e. the pristine environment) 
the magnetic field is tightly coupled to the primordial gas ($Z=0$).
This is true, however, only at the Jeans scale $R_{\rm J}$.
Here, let us estimate the drift velocity $v_{{\rm B}x}$ at a smaller scale. 
This should be compared with the typical turbulent velocity $u_{\rm turb}$ at this scale. 
The drift velocity is inversely proportional to the length scale (eq. \ref{eq:vBx});
$v_{{\rm B}x} \propto \alpha^{-1}$ at a scale $\alpha R_{\rm J}$ with $\alpha < 1$. 
For the spectrum of the turbulent velocity, we assume the power-law form, 
$u_{\rm turb}\propto R^{\vartheta}$. Here, the power index $\vartheta$ equals 
to $1/2$ for the Burgers turbulence and $1/3$ for the Kolmogorov turbulence, respectively.
From consideration that the turbulence is driven 
by the gravitational collapse motion at the Jeans scale, 
the typical turbulent velocity at this scale is roughly 
given by the free-fall velocity $u_{\rm ff}$. 
Thus the turbulent velocity at 
the scale $\alpha R_{\rm J}$ is 
\begin{equation}
u_{\rm turb} = \alpha ^{\vartheta} u_{\rm ff}.\label{eq:uturb}
 \end{equation}
Using these relations, the ratio is 
\begin{equation}
\frac{v_{{\rm B}x}}{u_{\rm turb}} \Big | _{\alpha R_{\rm J}} = \frac{v_{{\rm B}x}(R_{\rm J})}{u_{\rm ff} }\alpha^{-(\vartheta+1)}, 
\end{equation} 
and the dissipation condition at the scale $\alpha R_{\rm J}$ is given by
\begin{equation}
\frac{v_{{\rm B}x}(R_{\rm J})}{u_{\rm ff}} > \alpha^{\vartheta+1}.\label{eq:discon_vis}
\end{equation}
Note that the left hand side in the above inequality have been shown in
Fig. \ref{fig_cont_model1}d. 

On the other hand, 
the magnetic field grows fastest at the smallest scale of the
turbulence, the viscous scale $R_{\rm vis}$ defined by the relation
\begin{equation}
 u_{\rm turb}R_{\rm vis}=\nu_{\rm vis},\label{eq:vis_scale}
\end{equation}
where $\nu_{\rm vis}$ denotes the kinematic viscosity.
Combinig the relations (\ref{eq:uturb}) and (\ref{eq:vis_scale}) and using 
the Reynolds number ${\rm Re}\equiv  u_{\rm ff} R_{\rm J}/\nu_{\rm vis}$, 
the viscous scale can be given as
$R_{\rm vis}\simeq R_{\rm J} {\rm Re}^{-1/(\vartheta+1)}$, i.e., 
$\alpha= {\rm Re}^{-1/(\vartheta+1)}$ for the viscous scale. 
Note that the Reynolds number is approximately ${\rm Re}\sim 10^6 (n_{\rm H}/1{\rm
cm^{-3}})^{1/2}$ in the present calculation \citep[see also][]{schober12}.
Substituting the parameter $\alpha= {\rm Re}^{-1/(\vartheta+1)}$ at the
viscous scale into the inequality (\ref{eq:discon_vis}), we have the dissipation condition at the viscous scale:
\begin{equation}
\frac{v_{{\rm B}x}(R_{\rm J})}{u_{\rm ff}}  > {\rm Re}^{-1} \sim
 10^{-6}\left(\frac{n_{\rm H}}{1{\rm cm}^{-3}}\right)^{-1/2}.\label{eq:vis_dis}
\end{equation}
It is worth noting that this condition does not depend on the power
spectrum of the turbulence. 

For instance, at $\sim 1{\rm cm}^{-3}$ as assumed in 
\citet{schober12}, the magnetic field is tightly coupled to the gas
for $v_{{\rm B}x}(R_{\rm J})/u_{\rm ff} < 10^{-6}$.  
As seen in Fig. \ref{fig_cont_model1} d, 
this corresponds to $B \la 10^{-7}$G at $\sim 1{\rm cm}^{-3}$. 
Therefore, the magnetic field can be amplified up to $\sim 10^{-7}$G by 
the small-scale dynamo at this density. 
Note that this maximal field strength is roughly equals to
$0.1 B_{\rm cr}$, which is dynamically non-negligible.
Below $\la 10^{8}{\rm cm}^{-3}$, 
the boundary of the dissipative region (eq. \ref{eq:vis_dis})
stays at $B\sim 10^{-7}-10^{-6}$G, since the inverse of the Reynolds number 
decreases as the density increases (eq \ref{eq:vis_dis}), cancelling
the positive slope of the level curves (Fig. \ref{fig_cont_model1}d, the
contours just below $B_{\rm cr}$).
This means that, despite of the cloud contraction, the magnetic field in the core remains 
roughly at the constant level as a result of the dissipation. 
Eventually, at $\sim 10^8{\rm cm}^{-3}$, 
the inverse of the Reynolds number becomes $\sim 10^{-10}$, coinciding with
the level of the vertical contour and the dissipation condition of eq. (\ref{eq:vis_dis}) is satisfied.
Hence, the magnetic flux at the viscous scale dissipates from the gas in a dense
circumstance such as $\ga 10^8{\rm cm}^{-3}$.

In summary, the rapid amplification of the field at the viscous scale
up to the level of $10^{-7}-10^{-6}$ G seems to be plausible, 
if the turbulence assumed here is present in the cloud of $\la 10^8{\rm cm}^{-3}$.@
It is worth noting that the amplified level of the field strength could
be dynamically important.

A caveat about the small-scale dynamo amplification is that, 
the nature of turbulence at very small scales, including the inverse cascading,  
in cosmological minihalos has not yet been studied in detail. 
A future effort to tackle this problem would be rewarded. 
 
\bigskip
We appreciate fruitful discussions with Masahiro Machida, and the support by Ministry of
Education, Science, Sports and Culture, Grant-in-Aid for Scientific
Research (C22540295:HS, B25287040:KO).


\appendix
{
Here we present the expressions for the chemical heating/cooling rate 
in the present calculations, which is the same as in \citet{omukai00}. 
The cooling rate (per unit volume) 
associated with the H$_2$ formation/dissociation can be written by using 
the related reaction rates as:
\begin{eqnarray}
\rho \Lambda_{\rm chem,H_{\rm 2}} &=&
4.48{\rm eV} {n({\rm H_2})\left(n({\rm H})k_{\rm dis1} + n({\rm H_2})k_{\rm dis2}\right)}\nonumber\\
&&- \left( 0.2{\rm eV} + \frac{4.2{\rm eV} }{1+n_{\rm H}/n_{\rm cr}} \right) n({\rm H}) n_{\rm H} k_{\rm gr}\nonumber\\
&&- \frac{
  {3.53{\rm eV} }n({\rm H})n({\rm e})k_{\rm H^-} 
+ {1.83{\rm eV} }n({\rm H})n({\rm H^+})k_{\rm H^+_2} 
+ {4.48{\rm eV} }\left(n({\rm H})^3k_{\rm 3body1}+n({\rm H_2})n({\rm H})^2k_{\rm
	  3body2}\right)}
{\left(1+n_{\rm H}/n_{\rm cr}\right)}
\end{eqnarray}
where $n_{\rm cr}$ is the critical density for H$_2$ deexcitation given by equation (23) 
of \citet{omukai00}, and we have used 
the reaction rate coefficints of H$_2$ formation on the grain surfaces($k_{\rm
gr}$; reaction 23 in Omukai 2000), through H$^-$ process ($k_{\rm H^-}$; 8), H$^+_2$ process ($k_{\rm
H^+_2}$; 10) the three-body reactions ($k_{\rm 3body1}$ and $k_{\rm 3body2}$; 19 and 20), 
and the collisional dissociation rates ($k_{\rm dis1}$ and $k_{\rm dis2}$; 13 and 21). 
In the above, $n({\rm X})$ denotes the number density of the species ``X'', 
while $n_{\rm H}$ is the H nuclei number density.
The net cooling rate associated with the H ionization is 
\begin{equation}
\rho \Lambda_{\rm chem,H^+} = 13.6{\rm eV}\left(\frac{d n({\rm H^+})}{d
 t} + k_{\rm rec} n({\rm H^+})n({\rm e}) - k_{\rm CR, H} n({\rm H}) 
-k_{\rm CR, H_2} n({\rm H_2})) \right), 
\end{equation}
where the second term comes from the assumption that the photons emitted in the 
radiative recombination does not contribute to the heating and the third and forth terms 
from that the CR ionization via
\begin{eqnarray}
{\rm H     +   CR}    &\rightarrow&   {\rm H^+    +   e}\\
{\rm H_2    +   CR}   &\rightarrow&   {\rm H^+    +   H     +   e}, 
\end{eqnarray}
whose rate coefficients are $k_{\rm CR, H}$ and $k_{\rm CR, H_2}$, respectively,
does not contribute to the gas cooling. 
Note that we consider the heating associated with the CR ionization separately.
Similarly, the cooling rates associated with the He and He$^+$ ionization, respectively, 
are given by
\begin{eqnarray}
\rho \Lambda_{\rm chem, He^+} &=& 24.6{\rm eV}\left(\frac{d n({\rm He^+})}{d
 t} + k_{\rm rec, He^+} n({\rm He^+})n({\rm e}) - k_{\rm CR, He} n({\rm He}) \right).\nonumber\\
\rho \Lambda_{\rm chem, He^{++}} &=& 79.0{\rm eV}\left(\frac{d n({\rm
 He^{++}})}{d t} + k_{\rm rec, He^{++}} n({\rm He^{++}})n({\rm e})
						\right),
\end{eqnarray}
where $k_{\rm rec, He^+}$ is the rate coefficient for the radiative recombination of He$^+$ 
(reaction 4 in Omukai 2000) and
$k_{\rm CR, He}$ is the rate coefficient for the He ionization by CR:
\begin{equation}
{\rm He    +   CR}   \rightarrow   {{\rm He}^+   +   \rm{e}}.
\end{equation}
Summing up all the contributions above, we have the chemical cooling rate:
\begin{equation}
\rho \Lambda_{\rm chem} = \rho \Lambda_{\rm chem, H_2} + \rho \Lambda_{\rm chem, H} +
 \rho \Lambda_{\rm chem, H^+} + \rho \Lambda_{\rm chem, He^{++}}.
\end{equation}
In our calculation, heating/cooling associated with 
the H$_2$ formation/dissociation can be important,
while the recombination/ionization contribution has little significance 
owing to the small ionization degree.

For high enough density ($n_{\rm H} > 10^{13}{\rm cm^{-3}}$), we employ
the net change of the number density of H$^+$, He$^+$, He$^++$ and H$_2$
($\Delta n_{\rm X}$) times the latent heat divided by $\Delta t$ 
as the chemical cooling/heating rate per unit volume:
\begin{eqnarray}
\rho \Lambda_{\rm chem} = 
 13.6{\rm eV} \frac{d n({\rm H^+})}{dt}
 +24.6{\rm eV} \frac{d n({\rm He^+})}{dt}
 +79.0{\rm eV} \frac{d n({\rm He^{++}})}{dt}
 -4.48{\rm eV} \frac{d n({\rm H_2})}{dt}
\end{eqnarray}

}

\end{document}